\documentclass[usenatbib]{mnras}

\usepackage{newtxtext,newtxmath}

\usepackage[T1]{fontenc}

\DeclareRobustCommand{\VAN}[3]{#2}
\let\VANthebibliography\thebibliography
\def\thebibliography{\DeclareRobustCommand{\VAN}[3]{##3}\VANthebibliography}

\usepackage{graphicx}	
\graphicspath{{Images/}}
\usepackage{amsmath}	
\usepackage[table,xcdraw]{xcolor}    
\usepackage{subfig}
\usepackage{comment}
\usepackage[normalem]{ulem}
\usepackage{physics}
\usepackage{xcolor}

\definecolor{royalazure}{rgb}{0.0, 0.22, 0.66}
\definecolor{auburn}{rgb}{0.43, 0.21, 0.1}
\definecolor{bostonuniversityred}{rgb}{0.8, 0.0, 0.0}
\definecolor{planet}{HTML}{69DF45}

\usepackage{hyperref}
\hypersetup{
    colorlinks=true,
    linkcolor=bostonuniversityred,
    linktoc=all,
    filecolor=blue,      
    urlcolor=royalazure,
    citecolor=royalazure
}

\title[Gravitational instability in irradiated discs]{Short-Lived Gravitational Instability in Isolated Irradiated Discs} 

\author[S. Rowther et al.]{Sahl Rowther,$^{1,2}$\thanks{E-mail: sahl.rowther@leicester.ac.uk}
Daniel J. Price$^{2}$
, Christophe Pinte$^{2,3}$, Rebecca Nealon$^{4,5}$, Farzana Meru$^{4,5}$ and
\newauthor  Richard Alexander$^{1}$
\\ 
$^{1}$School of Physics and Astronomy, University of Leicester, Leicester LE1 7RH, UK\\
$^{2}$School of Physics and Astronomy, Monash University, Vic. 3800, Australia\\
$^{3}$Univ. Grenoble Alpes, CNRS, IPAG, F-38000 Grenoble, France\\
$^3$Centre for Exoplanets and Habitability, University of Warwick, Coventry CV4 7AL, UK\\
$^{4}$Department of Physics, University of Warwick, Coventry CV4 7AL, UK
}

\date{Accepted XXX. Received YYY; in original form ZZZ}

\pubyear{2024}

\begin{document}
\label{firstpage}
\pagerange{\pageref{firstpage}--\pageref{lastpage}}
\maketitle 

\begin{abstract}
Irradiation from the central star controls the temperature structure in protoplanetary discs. Yet simulations of gravitational instability typically use models of stellar irradiation with varying complexity, or ignore it altogether, assuming heat generated by spiral shocks is balanced by cooling, leading to a self-regulated state. In this paper, we perform simulations of irradiated, gravitationally unstable protoplanetary discs using 3D hydrodynamics coupled with live Monte-Carlo radiative transfer. We find that the resulting temperature profile is approximately constant in time, since the thermal effects of the star dominate. Hence, the disc cannot regulate gravitational instabilities by adjusting the temperatures in the disc. In a $0.1M_\odot$ disc, the disc instead adjusts by angular momentum transport induced by the spiral arms, leading to steadily decreasing surface density, and hence quenching of the instability. Thus, strong spiral arms caused by self-gravity would not persist for longer than ten thousand years in the absence of fresh infall, although weak spiral structures remain present over longer timescales. Using synthetic images at 1.3mm, we find that spirals formed in irradiated discs are challenging to detect. In higher mass discs, we find that fragmentation is likely because the dominant stellar irradiation overwhelms the stabilising influence of $P\mathrm{d}V$ work and shock heating in the spiral arms.
\end{abstract}




\section{Introduction}

A thin gas disc is unstable to axisymmetric perturbations when
\begin{equation}
    \label{eq:Toomre}
    Q(R) = \frac{c_{\text{s}} \kappa}{\pi G \Sigma} < 1,
\end{equation}
where $c_\text{s}$ is the sound speed, $\kappa \approx \Omega$ is the epicyclic frequency assumed to be equal to the orbital frequency, $\Sigma$ is the surface density \citep{1964Toomre} and $G$ is the Gravitational constant. The disc is gravitationally stable at warmer temperatures (higher $c_\text{s}$) or when the disc is less massive or more spread out (lower $\Sigma$). For $Q \lesssim 1.7$, the disc can become unstable to \emph{non}-axisymmetric perturbations to form spiral arms \citep{2007Durisen}. 

Complicating matters, neither $c_{\rm s}$ nor $\Sigma$ remain constant in response to the perturbation. \citet{2001Gammie} used shearing box simulations with cooling to show that the non-linear outcome of gravitational instability is either fragmentation or self-regulation, depending on $\beta$, the ratio of the local cooling timescale to the orbital time, such that
\begin{equation}
    \label{eq:tcool}
    t_{\rm cool} = \frac{\beta}{\Omega}.
\end{equation}
This assumes that discs cool from their warmer midplane to their upper layers.
Self-regulation occurs when heating from $P\mathrm{d}V$ work and shocks is balanced by cooling, leading to a thermostat where gravitational instability controls the temperature and the disc hovers around $Q\approx 1$, resulting in gravito-turbulence \citep{2003bRice,2004Lodato}. Fragmentation occurs when cooling is faster than heating ($\beta \approx 3$; \citealt{2001Gammie}). The exact criterion depends on a number of factors, such as the equation of state \citep{2005Rice}, and numerical resolution \citep{2010Meru,2011Lodato,2012Meru}. A constant value of $\beta$ implies that cooling is scale-free. This prescription has been used to investigate the angular momentum transport associated with a particular value of $\beta$ \citep{2001Gammie,2009Cossins}, how dust collects in the spiral arms \citep{2004Rice,2006Rice,2012Gibbons,2016Booth,2021Baehr,2022Baehr,2024Rowther}, and how planets interact with gravitationally unstable discs \citep{2011Baruteau,2015Malik,2020Rowther,2020bRowther,2022bRowther}. 

But can such a self-regulated state exist in nature? Recent observations with the Atacama Large Millimetre/Submillimetre Array (ALMA) have revealed a subset of apparently massive protoplanetary discs with spiral arms, including Elias 2-27 \citep{2016Perez,2017Meru,2018bForgan,2018Hall,2018bHuang,2021Veronesi}, IM Lupi and GM Aur \citep{2023Lodato}.

The problem with applying the $\beta$-cooling prescription to observed protoplanetary discs is that it ignores the radiation from the central star \citep{2016Kratter}.
Observations reveal thermally stratified discs passively irradiated by their central star \citep{1987Adams,1987Kenyon,Pinte2018,2021Law}, with surface layers \emph{hotter} than the midplane. It is unclear whether self-regulation is possible in this case, since one would infer `$\beta$-heating', rather than `$\beta$-cooling'.

Previous authors have investigated the evolution of gravitationally unstable discs using improved thermodynamics \citep{2002Boss,2004Mejia,2007Boley,2007Stamatellos,2009aForgan,2023SteimanCameron}. A general conclusion of these studies is that the evolution is sensitive to the thermodynamic model. Some included irradiation: \cite{2008Cai} modelled irradiation from an envelope heated by the central star, by irradiating the disc with a blackbody flux of a characteristic temperature.  \citet{2011Kratter} derived a critical temperature above which a disc would be `irradiation-dominated' and gravitoturbulence is impossible because the viscous heating is negligible compared to the irradiation. \citet{2010Meru} used 3D radiation hydrodynamics with flux-limited diffusion, and prescribed the stellar irradiation by fixing the temperature at the disc surface. \cite{2011Rice} modified the cooling prescription in \cite{2001Gammie} to include irradiation as an additional heating source. \cite{2013aForgan,2020Haworth,2020aCadman} modelled stellar irradiation as a power-law temperature profile to which the disc cools. \cite{2016Hall} included irradiation as an additional term in the cooling rate. A general result of these works is that irradiation weakens gravitational instabilities, and none found thermal stratification. However, using shearing box simulations \cite{2017Hirose} found that irradiation did not have a major impact on gravitoturbulence.

In this Paper, we use 3D hydrodynamics coupled with live Monte-Carlo radiative transfer to simulate irradiated gravitationally unstable discs. This method allows us to model the disc thermodynamics realistically.  \S\ref{sec:model} describes the simulations, \S\ref{sec:results} presents our results and comparison with $\beta$-cooling. We discuss in \S\ref{sec:Discussion}.

\section{Methods}
\label{sec:model}

We use \textsc{Phantom}, a smoothed particle hydrodynamics (SPH) code developed by \cite{2018Price} coupled with \textsc{mcfost}, a Monte-Carlo radiative transfer code developed by \cite{2006Pinte,2009Pinte}. 

\subsection{Fiducial Disc Setup}

We modelled the disc using $10^6$ SPH particles between $R_{\mathrm{in}} = 4$~au and $R_\mathrm{out} = 100$~au in an $0.1M_{\odot}$ disc around a $1M_{\odot}$ star, modelled as a sink particle with an accretion radius of $4$~au. We set up an initial surface density profile $\Sigma = \Sigma_{0}  \left ( {R}/{R_{\mathrm{in}}} \right)^{-0.6} f_{s}$,
where ${\Sigma_{0} = 1.97 \times 10^2\,\mathrm{g\ cm}^{-2}}$,  and ${f_{s} = 1-\sqrt{R_\mathrm{in}/R}}$ smooths the surface density at the inner boundary of the disc. The temperature in our simulations was set from {\sc mcfost}, but we set the initial vertical extent of the disc according to $H/R = 0.05 \left ( {R}/{R_{\rm in}} \right)^{0.25}$. The \cite{2010Cullen} switch detects any shocks that form and generates the correct shock dissipation depending on the proximity to the shock. Near the shock, the linear shock viscosity parameter $\alpha_{\text{AV}}\to 1$. Further away, $\alpha_{\text{AV}}$ decreases to a minimum value of 0.1. We set the quadratic artificial viscosity parameter $\beta_{\text{AV}}=2$  \cite[][]{2015Nealon,2018Price}. The disc is resolved with initial approximate mean smoothing length over disc scale height is ${<}h{>} / H = 0.2$.

\begin{figure*}
    \centering
    \includegraphics[width=\linewidth]{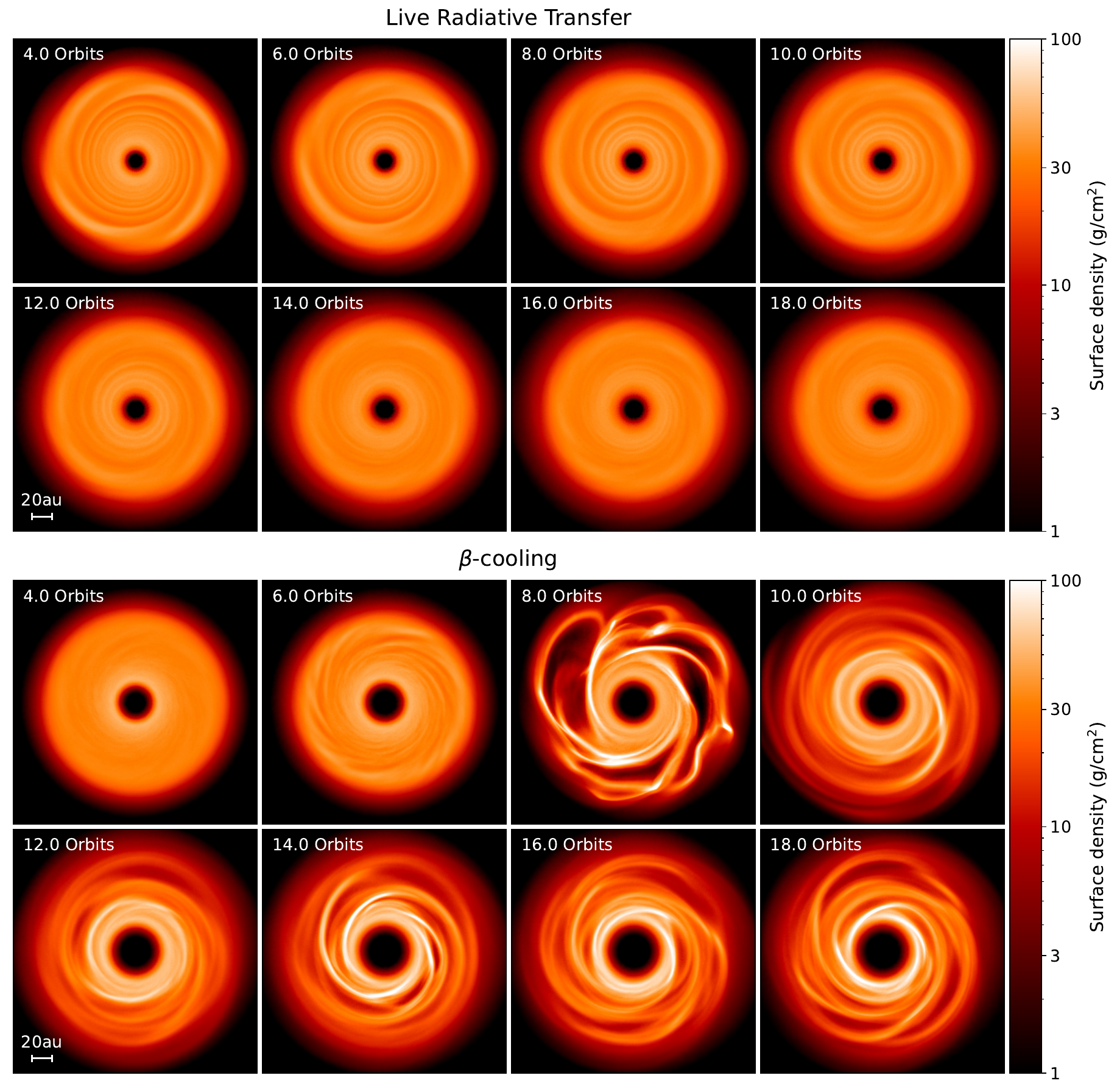}
    \caption{Surface density of a 0.1M$_\odot$ disc around a 1M$_\odot$, 0.88 L$_\odot$ star, shown every two orbits from 4 to 18 outer orbits. Each orbit is 1000 years. The top half shows the disc evolution using radiative transfer. Spiral arms develop due to gravitational instabilities within a couple of orbits. The disc remains strongly gravitationally unstable until around 10 outer orbits. However, this phase is short-lived. After ${\sim}$14 orbits, the spiral structures weaken and are barely noticeable. Lower half shows the disc evolution using $\beta$-cooling with no stellar luminosity. The spiral arms are stronger, and the disc remains gravitationally unstable.}
    \label{fig:densityEvol}
\end{figure*}

\begin{figure}
    \centering
    \includegraphics[width=\linewidth]{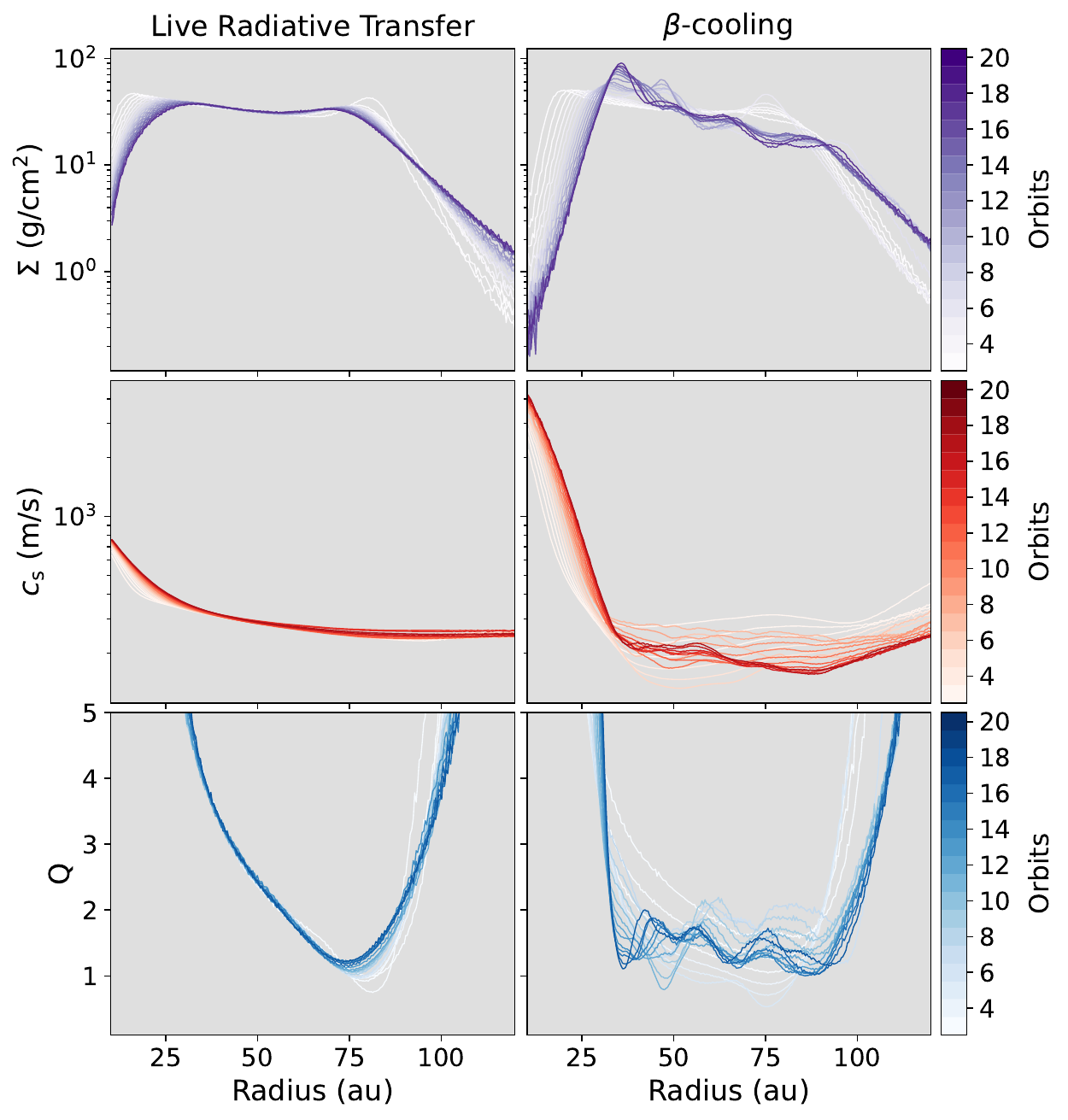}
    \caption{Azimuthally averaged plots of the surface density (top row), sound speed (middle), and Toomre $Q$ (bottom). We show azimuthally averaged quantities every outer orbit from 2 to 20 orbits. Darker lines represent later times. With $\beta$-cooling (right panels), the temperature of the disc evolves to compensate for changes in the surface density resulting in a steady gravitationally unstable state (see $Q$ profile). With radiative transfer (left panels) the temperature of the disc does not evolve. Since the dominant source of heating is due to stellar irradiation, which does not evolve in the timeframe of these simulations, the disc cannot cool down to compensate for the evolution of the disc surface density. Hence, as surface density decreases, $Q$ increases, resulting in a stable disc.
    }
    \label{fig:Q_plot}
\end{figure}

\subsection{Live Radiative Transfer}
\label{sec:LRT}

Temperatures in our simulation were calculated on-the-fly using  \textsc{mcfost}.
We included $P\mathrm{d}V$ work and shock heating as source terms for the radiative transfer, thus allowing self-regulation of the disc in principle since spiral shocks lead to heating.

We set the stellar luminosity from the sink particle mass and a 3Myr isochrone from \cite{2000Siess}. This corresponds to a stellar luminosity of $L=0.88L_\odot$, temperature of $T_\star = 4264$K, and a stellar radius of $R_\star = 1.72R_\odot$.
Since we simulated only gas, for radiative transfer we assumed perfectly coupled dust with a constant dust-to-gas mass ratio of 0.01. Dust in {\sc mcfost} was distributed between 0.03--1000$\mu$m  using 100 grain sizes and a power-law exponent of $\mathrm{d}n(s) \propto s^{-3.5} \mathrm{d}s$. In our disc these dust sizes gave Stokes numbers below 1, so the assumption that the dust is coupled to the gas is valid. We assumed spherical, homogenous, astrosilicate dust grains \citep{Weingartner2001}. Optical properties were calculated from Mie theory.

We illuminated the disc using $5 \times 10^9$ photons. Each \textsc{mcfost} Voronoi cell corresponded to one SPH particle. Our number of photons was larger than in previous studies with \textsc{Phantom + mcfost} \citep{2020Nealon,2022aBorchert,2022bBorchert}. A ratio of 100 photons for each SPH particle is usually enough to obtain the temperatures. We found more photons necessary to ensure at least some photons reach the densest part of the spiral arms in the disc. Otherwise, the particles hit the floor temperature of 2.73K, triggering gravitational collapse, becoming optically thicker and harder for photons to reach in subsequent timesteps. The resulting negative feedback loop led to artificial fragmentation. The number of photons chosen was adjusted until every cell received enough photons for the temperatures to converge such that the floor temperature was unnecessary.

We assumed radiative equilibrium; heating and cooling equilibrate on a shorter timescale than we simulate (see Discussion). We updated temperatures every ${1/(100\sqrt{2}) \approx 0.007071}$ of an orbit at $R_\mathrm{out}$, i.e. every 7.071 years. The high frequency ensured that the dynamical time was always longer than the time between temperature updates. In these simulations, the dynamical time in the inner-most regions ($R \sim 20$au) is roughly 12 times longer than the time between temperature updates. Hence, despite the energy of the particles not evolving between temperature updates, the particles always have the correct temperatures based on their current location since they have not had the time to undergo significant dynamical evolution.

\begin{figure}
    \centering
    \includegraphics[width=0.9\linewidth]{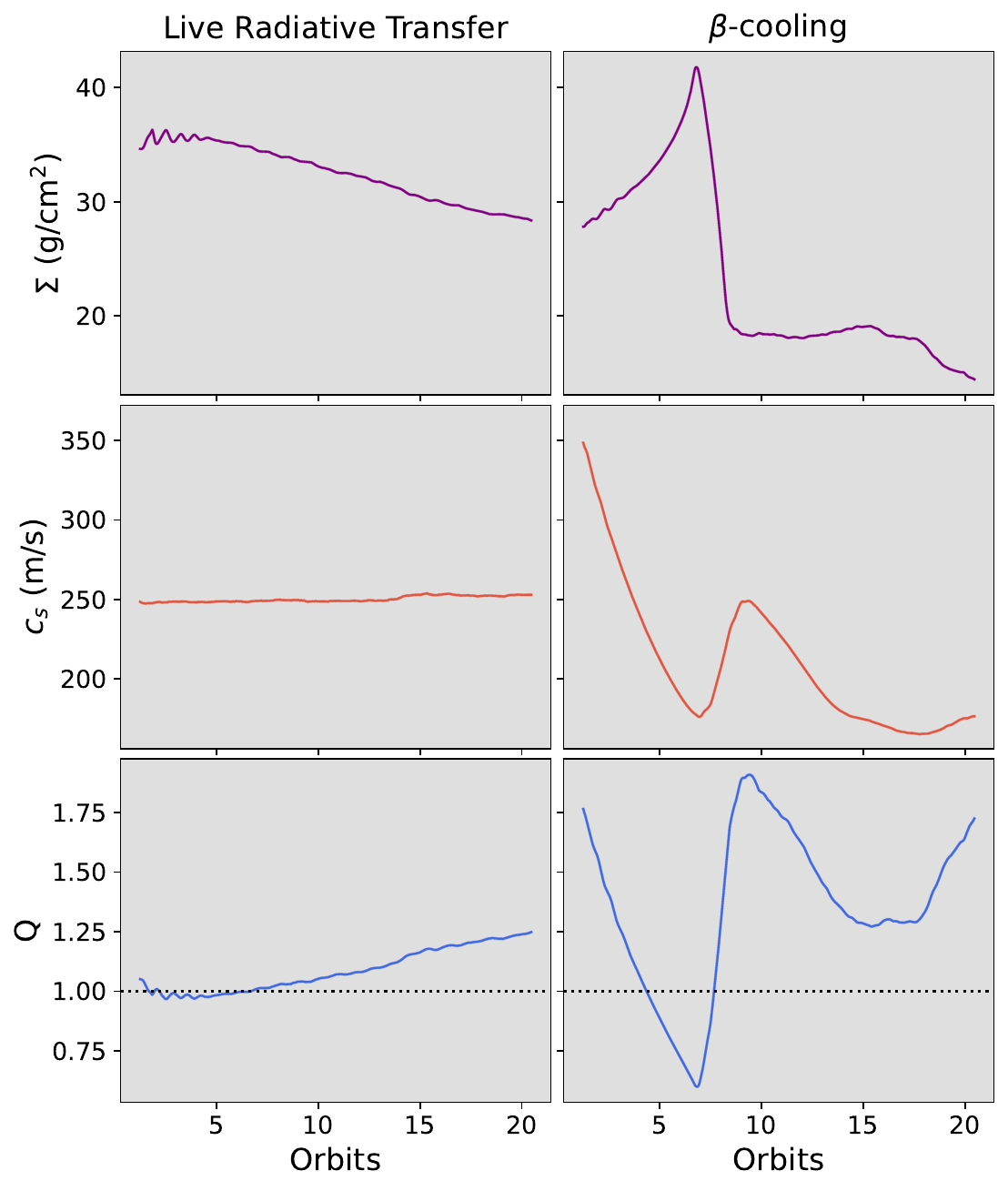}
    \caption{Time evolution of the surface density, sound speed, and Toomre $Q$ at $R = 77$~au (top, middle, and bottom panels, respectively). With $\beta$-cooling (right panels), the disc cools until it forms spiral arms. Once spiral shocks are strong enough (${\sim}8$ orbits), the disc then heats, eventually reaching a steady state after 14 orbits in terms of spiral structure, see Figure \ref{fig:densityEvol}. With radiative transfer (left panels), the disc steadily becomes more stable, as seen by the increasing $Q$. While the disc does heat up during the strongly gravitationally unstable phase (${\sim}2$--14 orbits) due to $P\mathrm{d}V$ work and shock heating by the spiral arms, the temperature change is tiny compared to the decreasing surface density that stabilises the disc. 
    }
    \label{fig:Q_R77}
\end{figure}

\subsection{Investigating Fragmentation}

We modelled a $0.25M_\odot$ disc to investigate whether fragmentation occurs when stellar irradiation is included. The disc was set up between $R_{\mathrm{in}} = 8$~au and $R_\mathrm{out} = 200$~au with an initial surface density profile of $\Sigma = \Sigma_{0}  \left ( {R}/{R_{\mathrm{in}}} \right)^{-1} f_{s}$, where $\Sigma_0 = 3.45 \times 10^2\,\text{g cm}^{-2}$. Additionally, for the live radiative transfer we used a 1Myr isochrone, and included accretion luminosity. All other parameters were the same as the fiducial simulation.

The changes from the fiducial simulation were numerically motivated by the need to avoid artificial fragmentation in transient features caused in the disc relaxation phase. These transients were common to both the live radiative transfer and $\beta$-cooling simulations. Physically, all the above changes to the setup favour stability by decreasing the amount of mass in the outer disc, and by using a more luminous star.

\subsection{$\beta$\,-\,Cooling}

For comparison to the fiducial simulation, we also simulated an identical disc cooled by $\beta$\,-\,cooling. The internal energy equation in this case is
given by
\begin{equation}
    \label{eq:enrg}
	\frac{\mathrm{d}u}{\mathrm{d}t} = -\frac{P}{\rho} \left ( \nabla \cdot \vb*{v} \right) + \Pi_{\mathrm{shock}} - \frac{u}{t_{\mathrm{cool}}},
\end{equation}
where we assume an adiabatic equation of state, $P$ is the pressure, $\rho$ is the density, and $u$ is the specific internal energy. The first term is the $P\mathrm{d}V$ work, $\Pi_{\mathrm{shock}}$ is the shock viscosity heating term, and we model $t_{\rm cool}$ using Eq~\ref{eq:tcool} with $\beta=15$.

\subsection{Synthetic Images}

We used \textsc{mcfost} to calculate synthetic continuum images at 1.3mm assuming the same stellar parameters as in \S\ref{sec:LRT}, and assuming a distance of 140pc.  The images were convolved using a 2D Guassian with a beam size of $0.05'' \times 0.05''$ (or $7\text{au} \times 7\text{au}$ at 140pc).

\section{Results}
\label{sec:results}

\subsection{Disc Evolution}

The top half of Figure \ref{fig:densityEvol} shows the surface density of the disc evolved with radiative transfer every two orbits from 4 to 18 outer orbits. Within four outer orbits, the disc develops spiral arms due to gravitational instabilities (top left panel). Spirals are strongest in the outer regions of the disc. This behaviour is consistent with expectations of realistic discs where gravitational instabilities are weaker in the inner warmer regions of the disc \citep{2005Rafikov,2009Stamatellos,2009Rice,2009Clarke}. For the next few orbits, there are no major changes to the morphology of the disc (middle and top right panels). After 12 orbits, the spiral arms are weaker (bottom left panel). As more time passes, the spiral arms continue to become weaker (middle and bottom right panels).

The weakening of gravitational instabilities is in contrast to the evolution of the disc with $\beta$-cooling. The bottom half of Figure~\ref{fig:densityEvol} shows that the $\beta$-cooled disc has reached a steady state with gravitational instabilities present throughout the disc. The density contrast of the spirals is larger with $\beta$-cooling compared to radiative transfer.

Figure~\ref{fig:Q_plot} compares the azimuthally averaged Toomre $Q$ parameter (bottom panels), sound speed (middle panels), and surface density (top panels) for the two simulations (left, radiative transfer; right, $\beta$-cooling). We plot azimuthally averaged quantities every outer orbit between 2 and 20 orbits.  Darker lines represent later times. 
Figure \ref{fig:Q_R77} shows the time evolution of surface density, sound speed, and $Q$ (top, middle, and bottom panels, respectively) at $R=77$~au (the location of minimum $Q$ at 10 outer orbits for the disc modelled using radiative transfer). The left and right panels represent the disc evolved with radiative transfer and $\beta$-cooling, respectively.

\begin{figure}
    \centering
    \includegraphics[width=\linewidth]{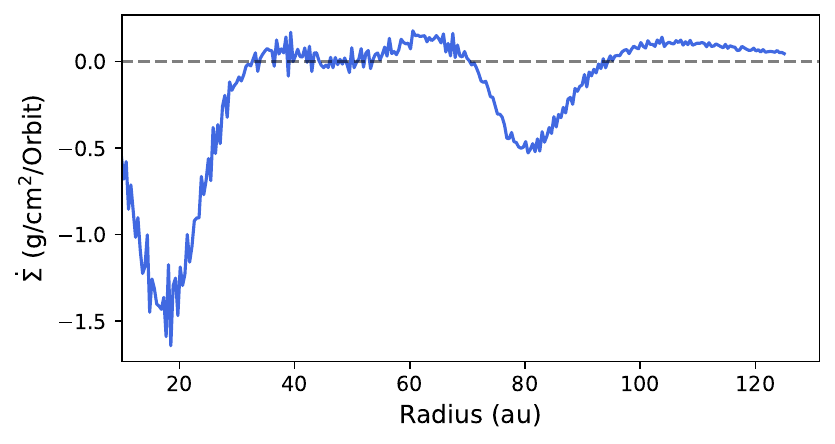}
    \caption{The rate of change of the surface density between 5 and 20 outer orbits at all parts of the disc for the radiative transfer simulation. The most gravitationally unstable parts of the disc lose the most amount of mass (at ${R = 80\pm10}$au). This mass loss results in the disc becoming more stable over time.}
    \label{fig:sigmadot}
\end{figure}

\begin{figure*}
    \centering
    \includegraphics[width=\linewidth]{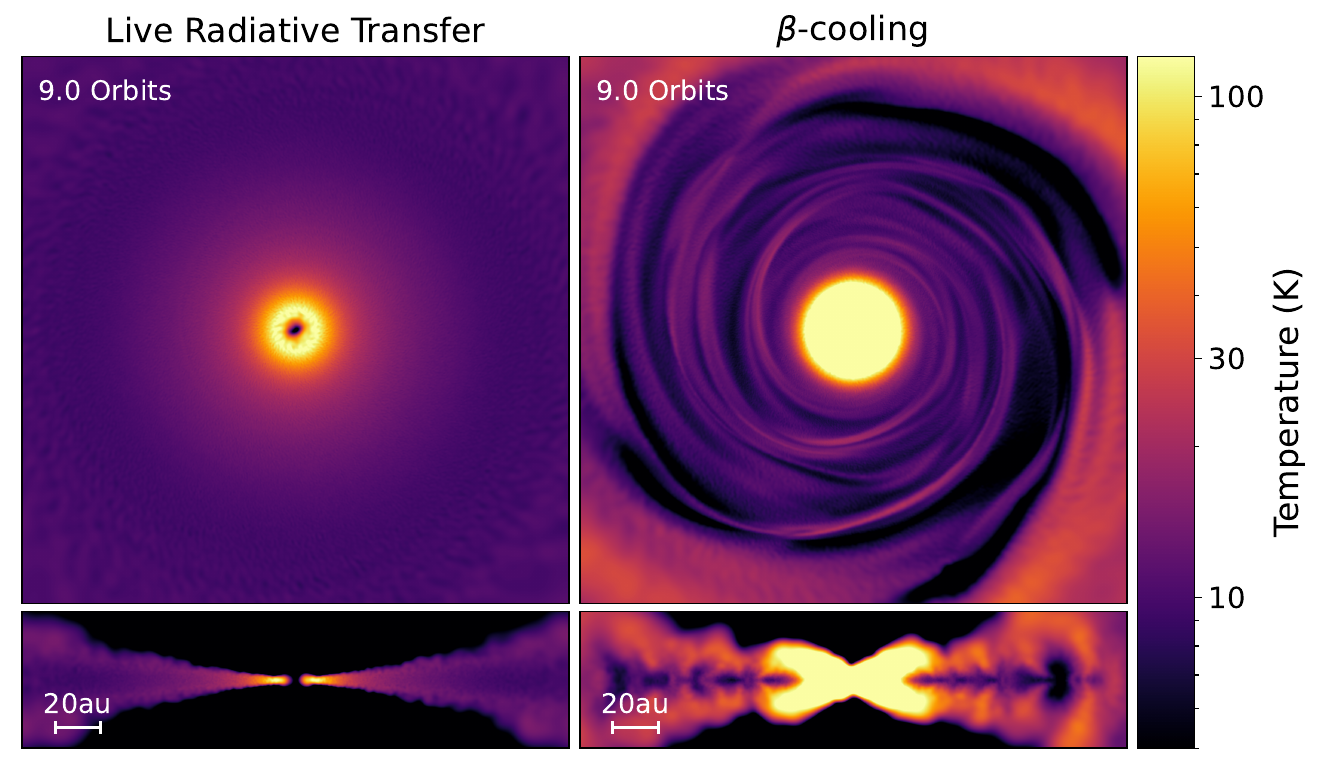}
    \caption{Cross-section slices of the face-on and edge-on views of the temperature structure of a $0.1M_\odot$ disc at 9 outer orbits. In contrast to the density structure, spirals are barely visible in temperature with irradiation (top left panel). $P\mathrm{d}V$ work and shock heating is negligible compared to the radiation of the star. Whereas with $\beta$-cooling, the temperature structure is strongly correlated to the density structure (top right panel) since the spirals are the only source of heating.  The irradiated disc is also thermally stratified and varies with radius, consistent with observations (bottom left panel). The vertical temperature structure with $\beta$-cooling (bottom right panel) is unusual due to the disc also spreading vertically. Thus, heating due to artificial viscosity is also relevant for the upper-most regions which is more poorly resolved. In the midplane, the disc temperature is uniform with radius unlike with radiative transfer. The bright 'x' shape in the inner-most regions is also due to a large inner cavity with very few particles.}
    \label{fig:temperature}
\end{figure*}

Figures \ref{fig:Q_plot} and \ref{fig:Q_R77} demonstrate why the evolution of the disc is different with radiative transfer compared to $\beta$-cooling.
With radiative transfer, the irradiation of the star dominates the temperature evolution. Thus, the temperature is fairly constant since the timeframe of these simulations is too short for the star to evolve. 
Whereas the surface density steadily decreases as seen in Figure \ref{fig:Q_R77}. The decrease in surface density is mirrored in the $Q$ profile by its steady increase. Hence, gravitational instabilities in the disc steadily weaken. In contrast, with $\beta$-cooling the evolution of the disc temperature is more dynamic.  From Figure \ref{fig:Q_R77}, the disc temperature continuously decreases before the onset of spiral structures since there is no source of heating. Once spirals form, $P\mathrm{d}V$ work and shock heating is able to balance the cooling resulting in a disc with gravitational instabilities.

Figure \ref{fig:sigmadot} shows the rate of change of the surface density between 5 and 20 outer orbits at all radial locations for the radiative transfer simulation. The regions that lose the most mass are also the most gravitationally unstable regions (${R = 80\pm10}$au). Hence, the disc becomes more stable as seen by $Q$ increasing in the outer parts of the disc in Figures \ref{fig:Q_plot} and \ref{fig:Q_R77}.

\subsection{Disc Temperatures} 

Figure \ref{fig:temperature} shows the temperature profiles of the discs modelled with {\sc phantom\,+\,mcfost} (left) and $\beta$-cooling (right). Top and bottom panels show face-on and edge-on views, respectively. Self-consistently calculating the temperatures results in two important differences compared to $\beta$-cooled simulations. Firstly, the temperature is controlled by the irradiation of the star rather than the balance between cooling, and $P\mathrm{d}V$ and shock heating. This is seen in the top left panel by the smooth temperature profile and the lack of spiral arms when the disc is modelled with radiative transfer. Whereas with $\beta$-cooling, the spirals are the only source of heating as seen by the spiral structure in the top right panel. Secondly, the temperature of the disc varies both radially and vertically with radiative transfer as seen in the bottom left panel. However, with $\beta$-cooling, the midplane temperature is radially constant (right panels). The unusual structures in the bottom right panel are due to poorly resolved regions in $\beta$-cooled simulations. The bright 'x' shape in the centre is due to a larger cavity in the inner disc. Additionally, unlike with the radiative transfer simulations, $\beta$-cooling also spreads the disc vertically resulting in the upper-most regions of the disc being more poorly resolved. Thus, heating due to artificial viscosity also becomes relevant in the upper parts of the disc with $\beta$-cooling.

\begin{figure*}
    \centering
    \includegraphics[width=\linewidth]{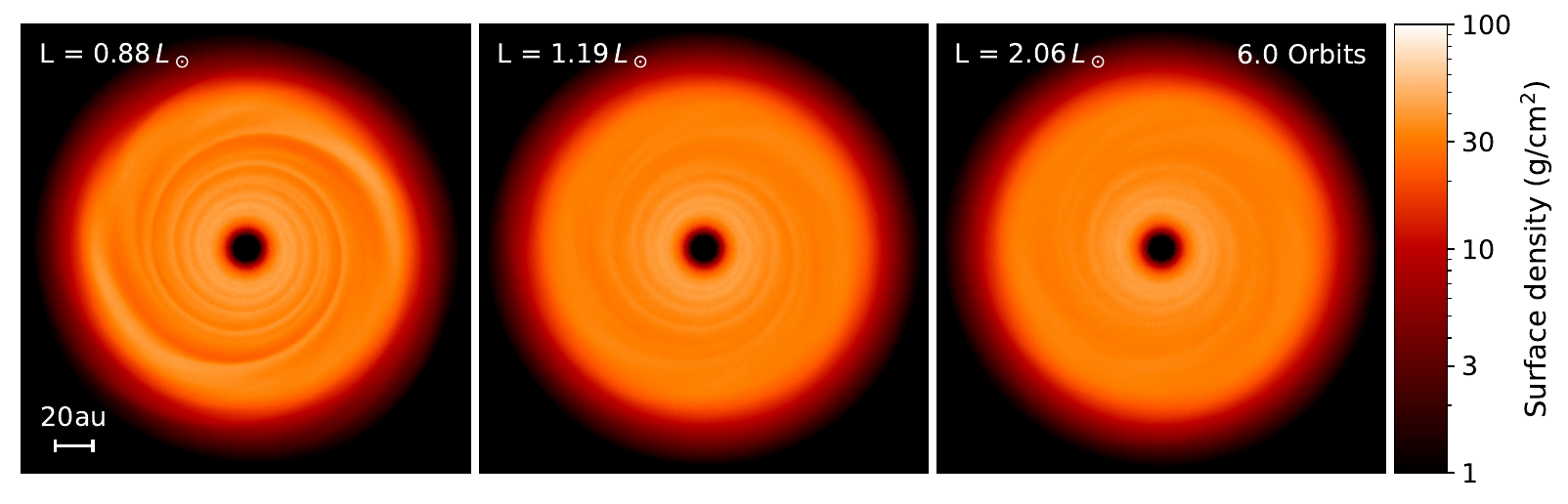}
    \caption{Surface density of a $0.1M_\odot$ disc after 6 orbits. From left to right, the luminosity of the star is 0.88, 1.19, and 2.06$\,L_\odot$ using a 3, 2, and 1\,Myr isochrone, respectively. Gravitational instabilities become weaker with increasing stellar luminosity.}
    \label{fig:star_comparison}
\end{figure*}

\subsection{Importance of the Star}

\begin{figure*}
    \centering
    \includegraphics[width=\linewidth]{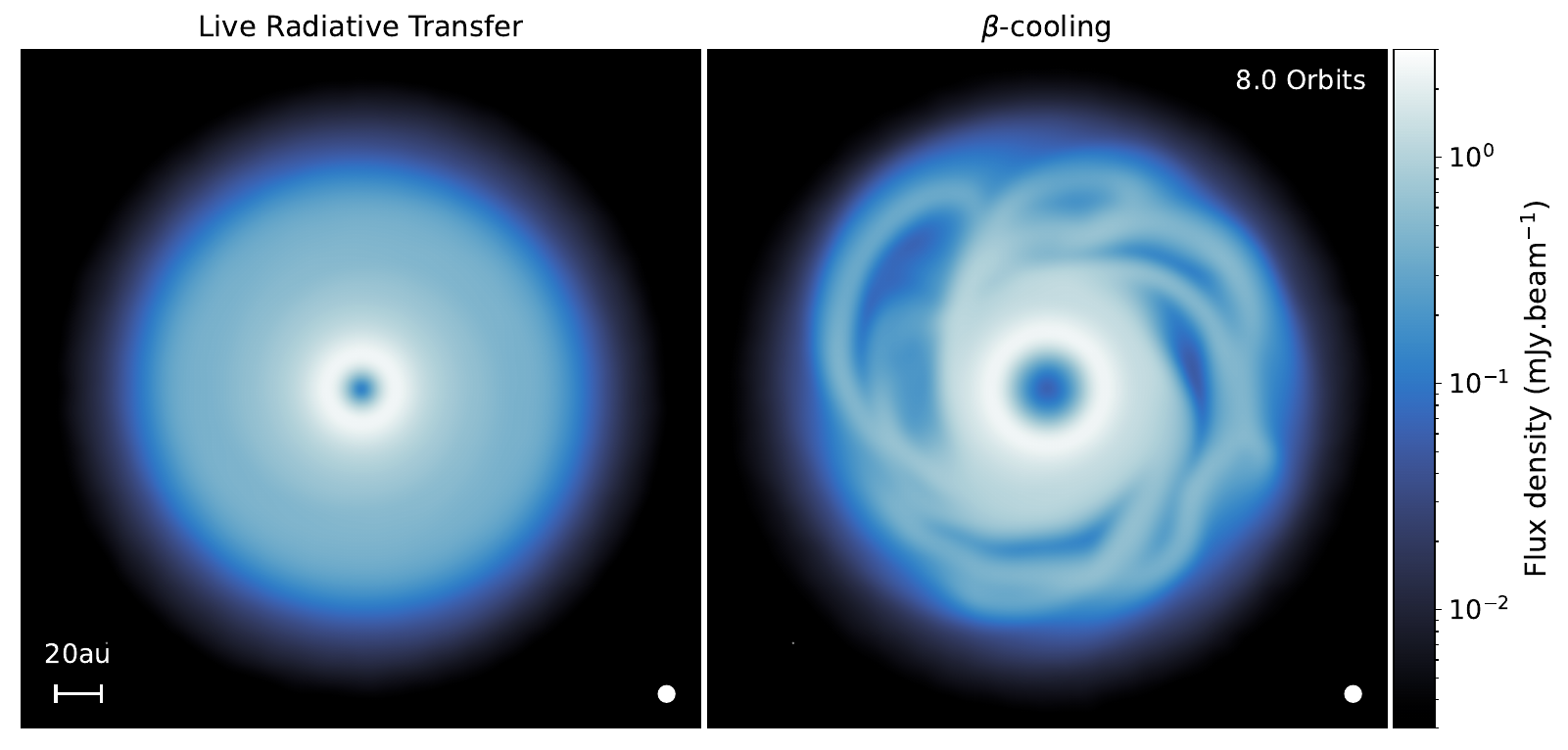}
    \caption{Synthetic images at 1.3mm of the disc simulated with radiative transfer (left) and with $\beta$-cooling (right). The beam size ($7\text{au} \times 7\text{au}$) is represented by the white circle. With radiative transfer, the spiral structures seen in the simulations would be challenging to detect in observations. Whereas with $\beta$-cooling, the spiral structures are much easier to detect.
    }
    \label{fig:synthImg}
\end{figure*}

Figure \ref{fig:star_comparison} compares the evolution of the disc modelled with radiative transfer using the 3 different isochrones to set the stellar luminosity. The reference simulation using a 3\,Myr isochrone with $L = 0.88L_\odot$, $T_\star = 4264$K, and $R_\star = 1.72R_\odot$ is shown in the left panel. The middle and right panels show the simulations using a 2\,Myr isochrone with $L = 1.19L_\odot$, $T_\star = 4277$K, and $R_\star = 1.99R_\odot$ and using a 1\,Myr isochrone with $L = 2.06L_\odot$, $T_\star = 4280$K, and $R_\star = 2.62R_\odot$, respectively. The stellar parameters impact the gravitational instabilities in the disc. The more luminous the star, the hotter the disc. Hence, gravitational instability becomes weaker with increasing stellar luminosity.

\subsection{Detectability of the Spirals}

Figure \ref{fig:synthImg} shows the synthetic images at 1.3mm of the disc simulated with radiative transfer (left) and with $\beta$-cooling (right) at 8 outer orbits. With radiative transfer the spirals are too weak to be detectable, even with high resolution, despite the spiral structures seen in the simulations (see Fig \ref{fig:densityEvol}). Whereas, with $\beta$-cooling, the spiral structures are apparent. The increased challenge in observing spirals when irradiation is included is consistent with \cite{2016Hall}. However, if dust were included in the simulations, concentration of solids in the spirals could make it easier to detect spiral structures \citep{2015bDipierro,2020Cadman}.

\subsection{Fragmentation in Higher Mass Discs}

\begin{figure}
    \centering
    \includegraphics[width=\linewidth]{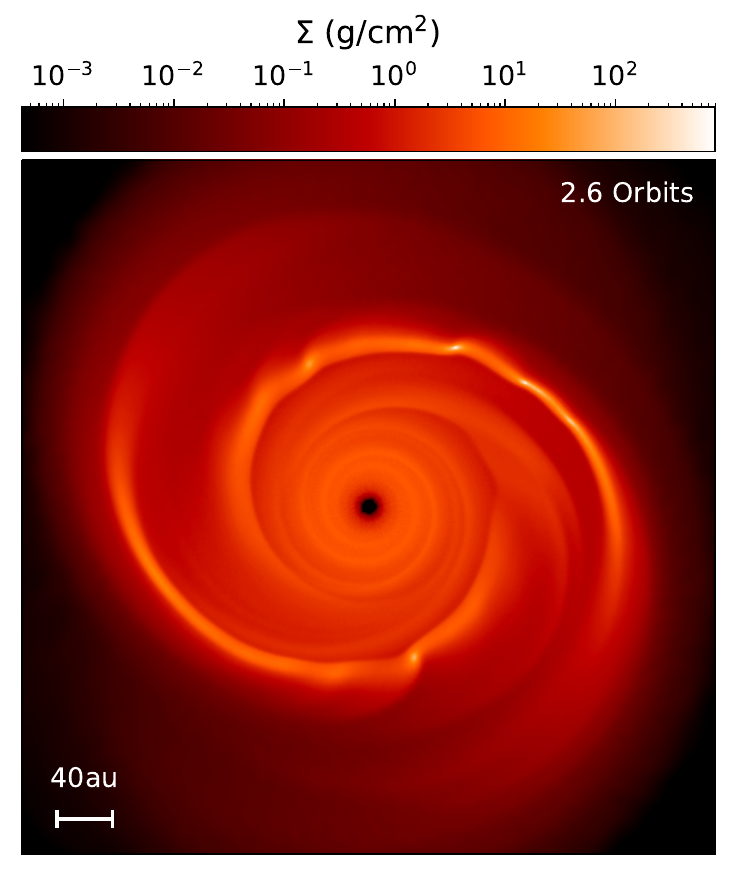}
    \caption{Fragmentation of a $0.25M_\odot$ disc modelled with radiative transfer. Since the irradiation of the star controls the disc temperatures, the disc is unable to stabilise and avoid fragmentation through $P\mathrm{d}V$ work and shock heating in the spiral arms.
    }
    \label{fig:Md0p25}
\end{figure}

Figure \ref{fig:Md0p25} shows the final state of a $0.25M_\odot$ disc modelled with radiative transfer. Unlike the fiducial simulation, the disc is now massive enough that the disc cannot avoid fragmentation. In a $\beta$-cooled disc, fragmentation could be prevented as the disc could regulate its temperatures through $P\mathrm{d}V$ work and shock heating in the spiral arms. However, this is negligible in an irradiated disc where the temperatures are set by the star since the thermal effects of stellar irradiation dominates. Hence, massive discs cannot stabilise through $P\mathrm{d}V$ work and shock heating, confirming the predictions in \cite{2011Kratter}. This is in contrast to previous simulations where stellar irradiation was modelled with a power-law which found fragmentation can be completely suppressed in discs irradiated by a $1M_\odot$ star. \citep{2020aCadman}.

\section{Discussion}
\label{sec:Discussion}

\subsection{A Different Fate}

With $\beta$-cooling, the disc cools until gravitational instability produces spiral arms. The spiral shocks heat the disc. As the disc heats up and becomes stable, $P\mathrm{d}V$ work and shock heating lessen. The constant $\beta$-cooling then cools and destabilises the disc. Eventually, an equilibrium state occurs where heating balances cooling represented by a consistent spiral structure.

Stellar irradiation changes the evolution and fate of a gravitationally unstable disc \citep{2011Kratter,2016Kratter}. For a lower mass gravitationally unstable disc, the equilibrium state is now determined by the irradiation of the star and by changes to the surface density of the disc. The star supplies a constant source of heat which dominates over other forms of heating, making the irradiated disc behave more like an isothermal simulation \citep{2016Kratter} where disc evolution is instead driven by changes in surface density. Angular momentum transport by the spiral arms results in viscous spreading which reduces the surface density and stabilises the disc over time. The strongly gravitationally unstable phase is relatively short-lived (around 10 outer orbits, or 10\,000 years). Weak gravitational instabilities can persist for longer. This makes observing the short gravitationally unstable phase difficult, at least in the absence of infall, consistent with the rarity of such discs in observations.

\cite{2020aCadman} found stellar irradiation suppressed fragmentation even for discs around solar mass stars even with disc-to-star mass ratios of 1. In contrast, we find fragmentation remains a possible fate for more massive discs (${>}\,0.25M_\odot$). The dominance of stellar irradiation prevents the disc from stabilising due to $P\mathrm{d}V$ work and shock heating in the spiral arms. Hence, even in more massive discs, spiral structures due to gravitational instabilities may be short-lived as clump (planet)-disc interactions will determine the morphology of the disc \citep{2020bRowther,2022bRowther}.

Although the above timescale for the spiral structures is similar to \cite{2019Hall}, a caveat to these results is that we do not model disc formation. In reality, discs will be more massive in the beginning and lose mass over time. Whether the initial larger disc masses lead to longer-lived spiral structures or rapid fragmentation is beyond the scope of this work.

\subsection{Self-Regulation by Infall?}
As outlined by \citet{2010aKratter}, the relevant thermal parameter is $\xi \equiv \dot{M}_{\rm in} G/ c_{\rm s}^3$, where $\dot{M}_{\rm in}$ is the mass infall rate from the environment. Thus while an irradiated disc cannot self-regulate by changing the sound speed, it may be possible to reach a steady state by fresh supply of mass \citep{2016Kratter}. \citet{2011Kratter} suggested that high mass, thick discs could transport sufficient angular momentum to avoid fragmentation so long as the mass infall remains below $\dot{M}_{\rm in} \sim c_{\rm s}^3/G$, potentially leading to steady mass accretion from excitation of global spiral modes \citep[see][]{1997Laughlin}. Examining if self-regulation could occur with mass infall in irradiation-dominated discs would make for worthwhile follow-up.

\subsection{Consequences for Planet Formation}

The stabilising nature of stellar irradiation has consequences for planet formation in gravitationally unstable discs. Planets forming via gravitational instability will be limited to only the most massive discs. Although gas fragmentation may be limited in lower mass discs by the weaker spiral structures, the dust could instead play a bigger role in forming planets in gravitationally unstable discs. Spiral arms are pressure maxima where dust can be trapped and grow to form planetesimals \citep{2004Rice,2006Rice,2012Gibbons,2016Booth,2020Elbakyan,2021Baehr,2022Baehr,2024Rowther}. The amplitude and number of spiral arms impact the dynamics of dust in the spiral arms. Dust becomes more excited in stronger spiral arms, or if they are more numerous \citep{2023Longarini,2023bLongarini}. Hence, weaker spiral arms with irradiation could be favourable to planetesimal formation as the dust could be less excited and collapse into clumps more easily. If planet formation occurs in these discs through gravitational collapse of gas or dust, then it must happen very early during the Class 0/1 phase when discs are massive enough to be gravitationally unstable \citep{2022Xu}.

The weaker spiral arms could also impact the fate of any planet that does form. \cite{2020Rowther,2020bRowther,2022bRowther} showed that thermodynamics strongly influences planet-disc interactions in gravitationally unstable discs. Weaker spiral arms driven by stellar irradiation could make it easier for a planet's spiral wakes to control the evolution of the disc by more easily suppressing the spirals due to gravitational instabilities. Thus, giant planets may have better survival odds than in $\beta$-cooled simulations where they rapidly migrate inwards \citep{2011Baruteau,2015Malik}. 

Protoplanetary discs do not evolve in isolation. The star-forming environment can shape the evolution of a disc through flybys or infall \citep{2018Bate}. Warping can also alter disc evolution \citep{2022Rowther}. In $\beta$-cooled discs, the disc returns to being unstable as it cools back down until it reaches a steady state with spiral shocks. With stellar irradiation, changes in surface density determine the fate of the disc. Flybys and infall change how mass is distributed. Thus, with stellar irradiation, any interactions with the star-forming environment that change the surface density profile of the disc could permanently alter the disc's evolution. 

\subsection{Importance of the Star}

Since stellar irradiation dominates the disc thermodynamics, the choice of stellar parameters is important for the evolution of the disc. The luminosities in this work ($0.88 - 2.06L_\odot$) are similar to observed stellar luminosities in discs thought to be gravitationally unstable such as Elias 2-27 and IM Lup \citep{2018bHuang} where the luminosity of the star is $0.91L_\odot$ and $2.57L_\odot$, respectively \citep{2018Andrews}.

\subsection{Validity of Assuming Radiative Equilibrium}

The main limitation of the radiative transfer simulations is the assumption of radiative equilibrium. In reality, discs do not radiate away their energy instantly. This assumption is valid if the timescale to reach temperature equilibration between the midplane and disc surface is sufficiently short. The timescale for temperature equilibration is given by $t_\text{rad} = \kappa \rho H^2/c$ \citep{1994Laughlin}, giving
\begin{equation}
    t_\text{rad} = 1.1\,\textrm{days} \left(\frac{\kappa}{5 \,\text{cm}^2/\text{g}}\right)  \left(\frac{\rho}{2\times10^{-13} \text{g}/\text{cm}^3} \right) \left(\frac{H}{3.0\,\textrm{au}}\right)^2,
\end{equation}
where we assume a scale height of 3 au (at R=77~au) to match our simulations. 

Compared to the dynamical time (676 years at 77~au) and the time between temperature updates (7.071 years), $t_\text{rad}$ (1.1 days) is insignificant. Therefore, we can approximate radiative transfer by assuming radiative equilibrium and keeping each particle's temperature fixed between temperature updates.

\section{Conclusions}

We use 3D hydrodynamics coupled with live Monte-Carlo radiative transfer to investigate the evolution of massive protoplanetary discs with irradiation from the central star. Our main findings are

\begin{enumerate}
    \item An irradiated gravitationally unstable disc does not regulate gravitational instabilities by adjusting the temperatures in the disc. The radiation of the star is the dominant source of energy, and is nearly constant over the timescale explored here (${\lesssim}\,100\,000$ years).
    \item In a lower mass gravitationally unstable disc ($0.1M_\odot$) the spiral structures are weaker compared to the spiral arms from {$\beta\text{-cooling}$}, consistent with previous works which included stellar irradiation, despite the differing methodologies \citep{2008Cai,2010Meru,2013aForgan}. The weaker gravitational instabilities are due to the minimum temperature of the disc set by the irradiation of the star. Unlike with $\beta$-cooling, the disc cannot continuously cool until strong spiral structures form.
    \item Fragmentation can occur in a higher mass gravitationally unstable disc ($0.25M_\odot$). Any additional heating from $P\mathrm{d}V$ work and shock heating in the spiral arms is negligible compared to the thermal effects of stellar irradiation. Hence, the disc cannot heat up and avoid fragmentation.
    \item In both scenarios, the spiral structures due to gravitational instabilities are short-lived. In lower mass discs, the evolution of the surface density is more important. The steady decrease in density due to angular momentum transport by the spiral arms results in the disc becoming more stable over time with weakening spiral structures. Whereas with higher mass discs, the disc quickly fragments resulting in a regime where clump-disc interactions will determine the morphology of the disc.
    \item Development of spiral arms remains possible in isolated irradiated discs, but is short-lived (${\sim}10\,000$ years) and thus an unlikely explanation for observed spiral structure in protoplanetary discs. The weaker spirals are also more challenging to detect.
\end{enumerate}

\section*{Acknowledgements}
DP thanks Kaitlin Kratter, Giuseppe Lodato and Alison Young for useful discussions. SR thanks Monash for hosting his visit. SR and RA acknowledge support funding from the Science \& Technology Facilities Council (STFC) through Consolidated Grant ST/W000857/1. FM acknowledges a Royal Society Dorothy Hodgkin Fellowship. RN acknowledges UKRI/EPSRC support via a Stephen Hawking Fellowship (EP/T017287/1). We acknowledge EU Horizon 2020 funding under the Marie Sk\l{}odowska-Curie grants 210021 and 823823 (DUSTBUSTERS). DP and CP acknowledge Australian Research Council funding via DP220103767 and DP240103290. We used the ALICE High Performance Computing Facility and the DiRAC Data Intensive service (DIaL), managed by the University of Leicester Research Computing Service on behalf of the STFC DiRAC HPC Facility (www.dirac.ac.uk). The DiRAC service at Leicester was funded by BEIS, UKRI and STFC capital funding and STFC operations grants. DiRAC is part of the UKRI Digital Research Infrastructure.

\section*{Data Availability}
Select snapshots of the simulations along with the data to recreate any of the simulations or the figures in the paper are available at \url{https://doi.org/10.5281/zenodo.13760732}. We utilised the following public software:\\

\begin{tabular}{ll}
\hspace{-0.6cm}\textsc{Phantom}
&\url{https://github.com/danieljprice/phantom}\\
& \citep{2018Price} \\
\hspace{-0.6cm}\textsc{mcfost} &\url{https://github.com/cpinte/mcfost}\\
& \citep{2006Pinte,2009Pinte} \\
\hspace{-0.6cm}\textsc{Splash}
&\url{https://github.com/danieljprice/splash}\\ & \citep{2007Price} \\
\hspace{-0.6cm}\textsc{Sarracen} &\url{https://github.com/ttricco/sarracen}\\
& \citep{sarracen}
\end{tabular}



\bibliographystyle{mnras.bst}
\bibliography{GI_Disc_RT.bib} 


\appendix




\bsp	
\label{lastpage}
\end{document}